\pdfoutput=1
\documentclass[reprint,amsmath,amssymb,aps,pra,superscriptaddress,floatfix]{revtex4-2}
\usepackage{graphicx}
\usepackage{booktabs}
\usepackage{bm}
\newcommand{\ad}{a^{\dagger}}
\newcommand{\nbar}{\bar{n}}
\newcommand{\Ep}{\hat{E}^{+}}
\newcommand{\Em}{\hat{E}^{-}}
\begin{document}
\title{Collective photon echoes in the Tavis--Cummings model:\\
distribution independence, two detuning regimes,\\
and the Dicke ladder}
\author{Michael T. Tavis}
\email{m.tavis@sbcglobal.net}
\affiliation{Retired, Lomita, California, USA}
\date{\today}
\begin{abstract}
An ensemble of $N$ two-level molecules prepared in its ground state and sharing a
lossless cavity with a weak field does not simply absorb: the intensity collapses and
then recurs, in a train of collective photon echoes. Working from the exact solution of
the Tavis--Cummings model in the few-photon regime $\nbar<N$, we confirm the echo time
$\tau_E=4\pi\sqrt{N+\Delta}/g$ (with $\Delta$ the squared scaled detuning,
Sec.~II) numerically at resonance from $N=5$ to $N=400$, and separate it from the
alternative $4\pi\sqrt{N-\nbar+\Delta}/g$ by a direct test: wherever a single
recurrence exists, its period is independent of $\nbar$ to the resolution of the
measurement ($0.8\%$) across a factor of six in $\nbar$, where the alternative
predicts shifts of up to $17\%$. We establish
four properties of the effect. The initial state requires no preparation, being the ground
state. The echo \emph{time} is independent of the initial photon distribution ---
coherent, thermal, sub- and super-Poissonian squeezed, and oscillatory squeezed
distributions spanning variances from $2$ to $34$ all recur together, as does a
controlled pair with identical mean and variance differing only in the shape of
$\rho_{nn}$ --- but the echo \emph{amplitude} is not: it varies at the tens-of-percent
level across distributions at fixed mean, including a factor of $1.8$ with the
\emph{squeezing phase} at fixed squeezing strength. Timing is distribution-blind;
amplitude carries distribution information. Detuning organizes the dynamics into two clean regimes separated by a
fragmented crossover, the echo time in the dispersive branch being exactly one-half that
of the resonant branch --- a factor we derive from second-order dispersive perturbation
theory and verify against the full dynamics, with the leading correction
$5Ng^{2}/\delta^{2}$ --- and sufficient detuning
removes the dependence on the initial Dicke state altogether. Finally, extending the treatment to arbitrary initial
$|r=N/2,m\rangle$, emission replaces absorption at $m\simeq-N/2+\nbar$, and the echo envelope
acquires one additional component for each step up the ladder, so that a
single-component echo occurs only at the ground state and the train blurs progressively
above it. A feasibility analysis against the measured parameters of an existing five-qubit
superconducting device --- including Lindblad simulations with cavity decay, qubit
relaxation, dephasing, frequency disorder and coupling inhomogeneity acting
\emph{simultaneously}, and full-Hilbert-space simulations of parameter disorder with the
permutation symmetry genuinely broken ---
shows the first echoes to be observable at $N\sim5$--$20$ on existing hardware: the echo
survives the dominant loss channel with contrast $\sim e^{-\kappa\tau_E/2}$, photons
being shielded from cavity decay while resident in the emitters.
\end{abstract}
\maketitle

\section{Introduction}

The exact solution of the Tavis--Cummings (TC) model \cite{TC1968} --- $N$ two-level
emitters (``molecules'' in the original terminology; transmons, atoms or spins in
current practice) coupled to a single quantized cavity mode --- has been available for
nearly six
decades, and the model itself is now realized routinely in circuit QED
\cite{Fink2009,Yang2020,Redchenko2025} and with trapped ions \cite{Ion2026}. Yet the
literature built on
that solution occupies a particular corner of it: the strong-field regime $\nbar\gg N$,
on resonance, with a coherent field, starting from the fully excited or fully unexcited
ensemble.

The closest existing study is that of Ramon, Brif and Mann \cite{Ramon1998}, who solved
the same invariant blocks numerically for initial Dicke states $|j,j\rangle$,
$|j,-j\rangle$ and $|j,0\rangle$ and established the collective modification of the
collapse--revival phenomenon, including a revival time
$\tau_R=2\pi\sqrt{\nbar+m+\tfrac12}$ and a halved revival at $m=0$. Their treatment is
resonant throughout, uses a coherent field throughout, and works at $\nbar=36\gg N$. The
revival there is governed by $\sqrt{\nbar}$.

A separate line of work has evaluated $\langle\Ep\Em\rangle(t)$ for a range of initial
photon distributions using the same exact reduction, but only at the endpoints of the
Dicke ladder --- the fully excited and fully unexcited ensembles
\cite{TC2013,Tavis2015} --- and identified persistent echoes in the few-photon regime
from the ground state \cite{Tavis2017}. The present work extends that line in three
directions: off resonance, to arbitrary $m$, and to nonclassical photon distributions;
and it establishes the governing scaling law together with the limits of its validity.

A recent treatment of superradiance initiated from different collective spin states
\cite{Alabbar2026} addresses a superficially similar question in a different setting:
free-space \emph{dissipative} emission treated by mean-field and Fokker--Planck methods,
with squeezing of the \emph{atomic} state. The present work concerns a lossless
single-mode cavity, exact finite-block dynamics, and squeezing of the \emph{injected
field}.

This paper concerns the opposite corner: $\nbar<N$, where the characteristic time is
governed by $\sqrt{N}$ instead, and where the initial state that matters experimentally
is the one requiring no preparation at all. We are not proposing a new method. The block
decomposition of the TC Hamiltonian, the collapse of the projection sum for a definite
Dicke state, and the $N^2$ scaling of the collective rate at mid-ladder are all
standard --- the last is the founding observation of the superabsorption and
quantum-battery literature \cite{Dicke1954,Higgins2014,Quach2022}. What is new is the
physics of the few-photon regime, which those bodies of work do not address: the echo
law and its verification across a factor of $80$ in $N$; the insensitivity of the echo
to the photon distribution, established including for oscillatory distributions at
matched moments; the two-regime structure of the detuning dependence; and the behaviour
along the Dicke ladder.

Section~\ref{sec:model} fixes the model and notation. Section~\ref{sec:echo} establishes
the echo law. Section~\ref{sec:dist} demonstrates distribution independence.
Section~\ref{sec:detune} gives the two detuning regimes. Section~\ref{sec:ladder} treats
general $m$. Section~\ref{sec:expt} is the feasibility analysis.

\section{Model, exact reduction, and conventions}
\label{sec:model}

We take the TC Hamiltonian in the rotating-wave approximation for a lossless cavity,
\begin{equation}
H=\hbar\omega_c\,\ad a+\hbar\omega_0 J_z+\hbar g\,(\ad J_-+a J_+),
\label{eq:H}
\end{equation}
with $g$ the single-molecule coupling. Throughout, times are quoted in the
dimensionless form $\gamma t$ with $\gamma\equiv g$, following Ref.~\cite{Tavis2017};
$\Ep\Em$ denotes the single-mode intensity operator, $\langle\Ep\Em\rangle
=\langle\ad a\rangle$ up to a constant prefactor that cancels in $S(t)$. In all figures
the sign of $S$ is chosen so that absorption is plotted positive for ground-state
initial conditions. The excitation number $c=n+m$ and the Casimir
$\mathbf{J}^2=r(r+1)$ are conserved, so $H$ is block diagonal; within a block the matrix
is tridiagonal with
\begin{equation}
|\langle n+1,m-1|\ad J_-|n,m\rangle|^{2}=(n+1)\big[r(r+1)-m(m-1)\big].
\label{eq:melem}
\end{equation}

A definite Dicke state with a definite photon number, $|n,m\rangle$, is a single basis
vector of the single block $(r,\,c=n+m)$. Consequently the projection of the initial
state onto the eigenbasis collapses to one eigenvector component, and the intensity
\begin{multline}
S(t)\equiv\langle\Ep\Em\rangle(t)-\langle\Ep\Em\rangle(0)\\
=-4\sum_{i<j}c_ic_j\,\langle E_i|\,\hat n-n\,|E_j\rangle\,
\sin^{2}\!\Big(\frac{\omega_{ij}t}{2}\Big),
\label{eq:S}
\end{multline}
with $c_i=\langle E_i|n,m\rangle$. This is an exact finite sum of sinusoids. This holds for \emph{every} $m$, not only the
endpoints; the point is elementary and is stated in Ref.~\cite{Ramon1998}, and we use it
as method rather than result. For an arbitrary distribution $\rho_n$ over photon number,
and more generally for any mixture over $(r,m)$, the result is an \emph{incoherent}
weighted sum, since $H$ and $\hat n$ are both block diagonal and cross-block coherences
cannot contribute to $\langle\Ep\Em\rangle$. We note the limitation that keeps this
honest: field-\emph{amplitude} observables such as $\langle\Ep\rangle$ connect
neighbouring blocks and admit no such reduction.

``Exact'' here means what it has always meant in this framework: exact reduction to
finite blocks, eigenvalues to arbitrary numerical precision, dynamics as an exact finite
sum. It does not mean closed-form radicals, which do not exist even at the endpoints for
blocks of dimension $>4$; the approximate closed forms available in certain limits were
developed in Ref.~\cite{TC1969}.

\paragraph{Detuning.} Within a block,
$\hbar\omega_c\ad a+\hbar\omega_0J_z=\hbar\omega_0 c+\hbar(\omega_c-\omega_0)n$; the
first term is constant across the block and cancels in every frequency difference.
Dividing by $\hbar g$ gives the dimensionless detuning
\begin{equation}
\beta=\frac{\omega_c-\omega_0}{g}=\frac{\delta}{g},
\qquad
\Delta\equiv\frac{\beta^{2}}{4}=\Big(\frac{\delta}{2g}\Big)^{2}.
\label{eq:beta}
\end{equation}
$S(t)$ is exactly even in $\beta$: the unitary $(-1)^{\hat n}$ maps the block
Hamiltonian at $-\beta$ to minus the Hamiltonian at $+\beta$, leaving every
$|\omega_{ij}|$ and every relevant matrix element invariant. Blue and red detuning are
therefore equivalent for this observable.

\paragraph{Numerical method and verification.} All results below were computed by two
independent implementations: the characteristic-polynomial construction of
Ref.~\cite{TC1968}, in the general-$(r,c)$ form given in Ref.~\cite{Tavis1968thesis},
evaluated at arbitrary precision, and an independent
machine-precision code building the block directly from Eq.~\eqref{eq:melem} and
diagonalizing it. The two agree on every peak-to-peak amplitude and recurrence
estimate reported (e.g.\ the $N=5$ amplitude $0.9777$ to four figures, the $N=50$
recurrence estimates to five). Sampling convergence was enforced by a physical ceiling: at resonance from the
ground state the peak-to-peak of $S$ cannot exceed $\nbar$, and undersampling clips it;
all traces reported reach that ceiling to better than $0.5\%$.

\section{The collective echo at resonance}
\label{sec:echo}

\begin{figure*}[!tb]
\centering
\includegraphics[width=\textwidth]{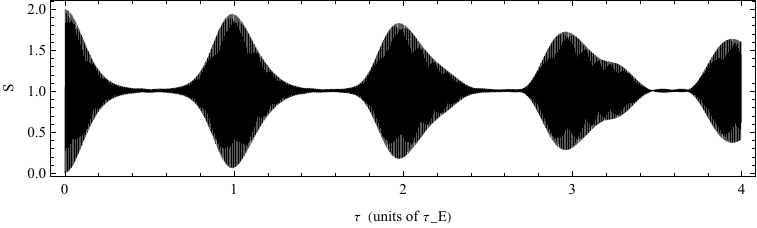}\\[1pt]
{\small (a)}\\[8pt]
\includegraphics[width=\textwidth]{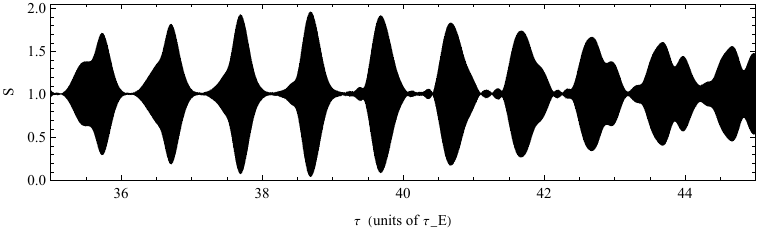}\\[1pt]
{\small (b)}
\caption{Collective photon echo.
$S(t)\equiv\langle\Ep\Em\rangle(t)-\langle\Ep\Em\rangle(0)$ for $N=50$ two-level
molecules initially in the ground state, sharing a lossless cavity with a coherent field
of mean photon number $\nbar=2$, on resonance. The horizontal axis is normalized by
$4\pi\sqrt{N-\nbar}/g=16\sqrt{3}\,\pi/g\simeq87.06/g$, the plotting convention of the
original machinery; Eq.~\eqref{eq:echo} gives $4\pi\sqrt{N}/g\simeq88.86/g$ for the same
quantity, a difference of $2.0\%$ that is below the resolution of the estimator and
shifts no feature visibly (see text).
(a) The echo train at early times: echoes at $\tau=1,2,3,4$; measured recurrence
$0.997\,\tau_E$.
(b) The same system at $\tau=35$--$45$, plotted identically. The echoes remain at unit
spacing and retain full contrast forty revival times later --- the individual maxima sit
near $\tau\simeq k+0.7$, a slow accumulated phase drift of the envelope, while the
spacing stays $1.00$ --- and their amplitude is modulated on a much longer scale,
peaking near $\tau\simeq40$. Over $\tau=0$--$200$ the re-emergence recurs at $\tau\simeq40,78,118,158$, agreeing with
$\tau_{\rm re}=\pi N\tau_E/4=39.3\,\tau_E$ of Eq.~\eqref{eq:reemerge}.
$S$ is plotted absorption-positive. The peak-to-peak value reaches $1.995$ against the
physical ceiling of $\nbar=2$, confirming numerical convergence.}
\label{fig:echo}
\end{figure*}

\subsection{The echo law and its verification in $N$}
\label{sec:echolaw}

For the ensemble prepared with all molecules in the ground state and a weak field,
$\nbar<N$, the intensity collapses and recurs at
\begin{equation}
\boxed{\;\tau_E=\frac{4\pi}{g}\sqrt{N+\Delta}\;}
\label{eq:echo}
\end{equation}
as established in Ref.~\cite{Tavis2017}. This is to be contrasted with
$2\pi\sqrt{\nbar+1+\Delta}/g$, the revival time for few molecules and a \emph{large} mean
photon number: the collective few-photon echo is governed by $\sqrt{N}$ where the
familiar case is governed by $\sqrt{\nbar}$.

\emph{Estimator.} Unless stated otherwise, a quoted recurrence is the dominant period of
the Fourier spectrum of the smoothed envelope of $S$ over a $20\,\tau_E$ window
(resolution $\sim5\%$); positions of individual envelope maxima are quoted where
explicitly identified as such. Over longer windows the envelope spectrum resolves fine
structure (Sec.~VI), and we do not mix the two.

\emph{Which form of the law?} Two candidate forms must be separated:
Eq.~\eqref{eq:echo} and the alternative $4\pi\sqrt{N-\nbar+\Delta}/g$, which the
computational machinery used as its normalization before the question was settled. The
two differ by $\nbar/2N$ to leading order --- $2\%$ at $N=50$, $\nbar=2$, below
resolution --- so only the small-$N$, larger-$\nbar$ corner discriminates between them.
At $N=5$, $\nbar=1$ they differ by $12\%$ ($28.1$ vs $25.1\,\gamma t$) and the
measurement decides: the dominant recurrence is $27.8\,\gamma t$ on the
$20\,\tau_E$-window estimator above ($\sim5\%$ resolution), and $27.4\,\gamma t$ on
the finer estimator of Sec.~\ref{sec:grid} ($\pm0.8\%$) --- $1$--$2.4\%$ from
$4\pi\sqrt{N}$ and $9$--$11\%$ from the alternative on either. Section~\ref{sec:grid}
makes the discrimination systematic by sweeping $\nbar$ at fixed $N$. The $\Delta$-dependence of Eq.~\eqref{eq:echo} is
supported by the Appendix derivation and by the two-regime structure of Sec.~V, but has
not been tested point-by-point within the resonant branch, where $\Delta\leq1$ shifts
$\tau_E$ by less than the resolution.

\begin{figure*}[!tb]
\centering
\includegraphics[width=\textwidth]{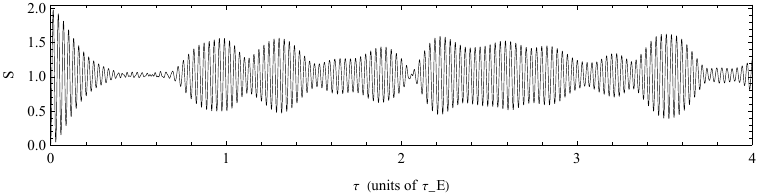}\\[-1pt]{\small (a) $N=10$}\\[4pt]
\includegraphics[width=\textwidth]{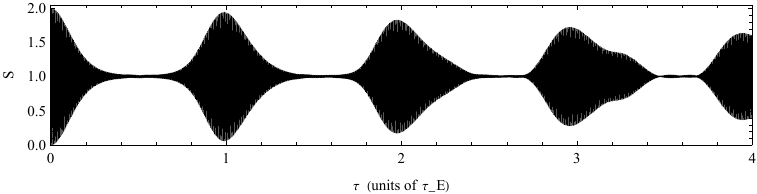}\\[-1pt]{\small (b) $N=50$}\\[4pt]
\includegraphics[width=\textwidth]{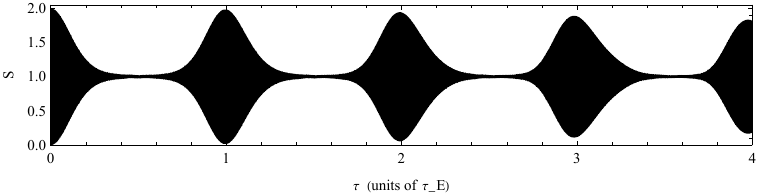}\\[-1pt]{\small (c) $N=100$}\\[4pt]
\includegraphics[width=\textwidth]{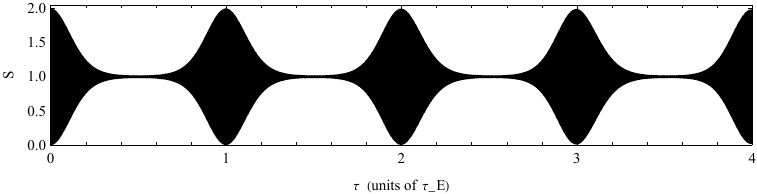}\\[-1pt]{\small (d) $N=400$}
\caption{Verification of the echo law across a factor of forty in $N$. Ground-state
ensemble, coherent field with $\nbar=2$, on resonance, in each case plotted against
$\tau=t/\tau_E^{\rm plot}$ with the normalization $\tau_E^{\rm plot}=4\pi\sqrt{N-\nbar}/g$
evaluated separately for that $N$: $35.5$, $87.1$, $124.4$ and $250.7$ in units of
$1/g$. On Eq.~\eqref{eq:echo} the corresponding values are $39.7$, $88.9$, $125.7$ and
$251.3$; the two agree to $2\%$ or better at $N\geq50$ and to $12\%$ at $N=10$, where
the train is in any case too ragged to time at that precision (the discrimination
between the forms is made in Sec.~\ref{sec:grid}, not here). Echoes occur at integer
$\tau$ throughout, confirming Eq.~\eqref{eq:echo}. The train sharpens systematically with $N$ ---
visibly ragged at $N=10$, where the recurrences are smeared into broad bursts, and
textbook at $N=400$ --- as expected of an asymptotic law. The experimentally relevant
range $N\sim5$--$50$ lies at the ragged end. All panels share the vertical scale, and
each reaches the convergence ceiling $S_{\rm pp}=\nbar=2$.}
\label{fig:acrossN}
\end{figure*}

\emph{A note on figure axes.} The traces in Figs.~\ref{fig:echo}, \ref{fig:acrossN} and
\ref{fig:dist}--\ref{fig:ladder} were
generated with the horizontal axis normalized by $4\pi\sqrt{N-\nbar+\Delta}/g$, the
convention of the original machinery. We have not regenerated them on the corrected
normalization, because at the parameters plotted the two differ by $2\%$ or less --- at
$N=50$, $\nbar=2$, $87.06/g$ against $88.86/g$ --- which is below the resolution of
every estimator used and shifts no feature by a visible amount. Each caption states the
numerical value of the normalizing constant actually used, so that the conversion is
explicit. All \emph{quantitative} statements in the text, every entry of
Table~\ref{tab:feas}, and the analysis of Sec.~\ref{sec:grid} use Eq.~\eqref{eq:echo}.

Equation~\eqref{eq:echo} is asymptotic in $N$, and we verify it accordingly.
Table~\ref{tab:echoN} and Fig.~\ref{fig:acrossN} give the measured recurrence from
$N=10$ to $400$ at $\nbar=2$ on resonance. The echoes sit at integer $\tau$ throughout,
and the \emph{sharpness} improves with $N$: at $N=10$ the train is visibly ragged, at
$N=400$ it is textbook. At $N=50$ the spectral estimator gives $0.997\,\tau_E$; over
much longer windows the envelope spectrum resolves fine structure near $83$ and
$85\,\gamma t$ (Sec.~VI), which we report as structure rather than as a competing single
number.

\begin{table}[h]
\centering
\begin{tabular}{cccc}
\toprule
$N$ & $4\pi\sqrt{N}$ & $4\pi\sqrt{N-\nbar}$ & echoes observed at\\
\midrule
10  & 39.7  & 35.5  & integer $\tau$ (ragged)\\
50  & 88.9  & 87.1  & integer $\tau$\\
100 & 125.7 & 124.4 & integer $\tau$\\
400 & 251.3 & 250.7 & integer $\tau$ (sharp)\\
\bottomrule
\end{tabular}
\caption{Verification of Eq.~\eqref{eq:echo} at $\nbar=2$, resonance. Times in units of
$\gamma t$. The second column is the law of Eq.~\eqref{eq:echo}; the third is the
alternative form used as the plotting normalization (see text), shown for comparison.
At $N\geq50$ the two are indistinguishable at the resolution of the estimator; the
discrimination between them is made in Sec.~\ref{sec:grid}.}
\label{tab:echoN}
\end{table}

The experimentally relevant range, $N\sim5$--$50$, therefore sits at the ragged end of an
asymptotic law. We state this explicitly rather than quoting the large-$N$ behaviour
alone.

\subsection{Which law: a direct test of the $\nbar$ dependence}
\label{sec:grid}

The verification above fixes $\nbar=2$ and varies $N$. Since the two candidate forms
differ precisely in their $\nbar$ dependence --- Eq.~\eqref{eq:echo} predicts no
dependence at all, the alternative predicts a period falling as $\sqrt{N-\nbar}$ ---
the decisive test is to hold $N$ fixed and sweep $\nbar$. We do this with an estimator
that makes no reference to either law, and we pay particular attention to $N=5$: it is
where the two laws differ most ($12\%$ at $\nbar=1$, against $2\%$ at $N=50$), and it
is the size of the benchmark device of Sec.~\ref{sec:expt}.

\emph{Method.} For each $(N,\nbar)$ the envelope of $S(t)$ is taken over $20$ nominal
periods and its spectrum examined in a broad band, $0.4$ to $2.5$ times
$4\pi\sqrt{N}/g$, wide enough to contain either candidate and their neighbours. We
record the dominant period and the amplitude of the next distinct component relative
to it. A point is said to possess a single recurrence only when that ratio is below
$0.5$; otherwise the envelope carries several components of comparable weight, no
single period is defined, and the point is reported as such rather than fitted. The
spectral resolution, checked by resampling, is $\pm0.8\%$ in period.

\begin{figure*}[!tb]
\centering
\includegraphics[width=\textwidth]{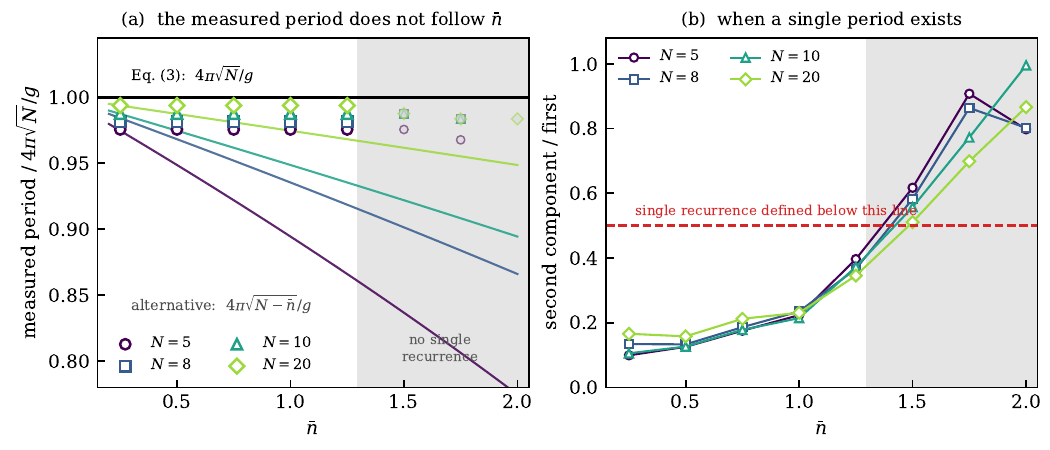}
\caption{Direct test of the $\nbar$ dependence of the echo period. Ground state,
resonance, coherent field; the envelope spectrum of $S(t)$ over $20$ periods, with no
reference to either candidate law.
(a) Measured dominant period, normalized to $4\pi\sqrt{N}/g$, against $\nbar$ for
$N=5$, $8$, $10$ and $20$. Open symbols are points possessing a single recurrence;
faded symbols lie in the multi-component region of panel (b). The solid black line is
Eq.~\eqref{eq:echo}; the coloured curves are the alternative $4\pi\sqrt{N-\nbar}/g$ for
each $N$. The measured period is flat in $\nbar$ to the $\pm0.8\%$ resolution --- at
$N=5$ it is $27.41/g$ at every $\nbar$ from $0.25$ to $1.5$ --- while the alternative
falls by $17\%$, $10\%$, $8\%$ and $4\%$ over $\nbar=0.25$ to $1.75$.
(b) Ratio of the second envelope component to the first. A single recurrence is defined
below the dashed line; the ratio crosses $0.5$ at $\nbar\simeq1.4$ for every $N$
examined, above which the echo envelope carries several rhythms of comparable weight
(shaded).}
\label{fig:grid}
\end{figure*}

\emph{Result.} Figure~\ref{fig:grid}(a) shows the sweep. At $N=5$ the measured period is
$27.41/g$ at $\nbar=0.25$, $0.5$, $0.75$, $1.0$, $1.25$ and $1.5$, and $27.19/g$ at
$1.75$ --- constant to a single resolution step across a factor of seven in $\nbar$ ---
where the alternative law predicts a fall from $27.39$ to $22.65$, i.e.\ by $17\%$. The
same holds at $N=8$ ($34.89$--$35.10/g$; alternative falls $10\%$), at $N=10$
($39.25/g$ at every $\nbar\leq1.5$; $8\%$) and at $N=20$ ($55.85\to55.28$; $4\%$). The measured period does not
follow $\nbar$. This is the property Eq.~\eqref{eq:echo} asserts and the alternative
denies, and it is measured directly rather than inferred from a fit.

Comparing the absolute value with Eq.~\eqref{eq:echo} over the $14$ grid points
possessing a single recurrence ($N=5,8,10,15,20,30,50$ at $\nbar=0.5$ and $1$):

\begin{table}[h]
\centering
\begin{tabular}{lrr}
\toprule
 & mean $|$deviation$|$ & worst case\\
\midrule
$\tau_E=4\pi\sqrt{N}/g$        & $1.1\%$ & $2.4\%$\\
$\tau_E=4\pi\sqrt{N-\nbar}/g$  & $2.3\%$ & $9.1\%$\\
\bottomrule
\end{tabular}
\caption{Accuracy of the two candidate laws over the grid points at which a single
recurrence exists. Resonance, ground state, coherent field; resolution $\pm0.8\%$.}
\label{tab:grid}
\end{table}

\noindent The deviation from $4\pi\sqrt N/g$ is $-2.4\%$ at $N=5$, $-1.8\%$ at $N=8$,
$-1.2\%$ at $N=10$, $-0.6$ to $-1.2\%$ at $N=15$--$30$ and $0.0\%$ at $N=50$: small, negative,
and shrinking monotonically with $N$, as an asymptotic law should. The alternative's
worst case, $9.1\%$, occurs at $N=5$, $\nbar=1$, where the two laws are most
distinguishable.

\emph{Where a single period ceases to exist.} Figure~\ref{fig:grid}(b) shows the
second-to-first component ratio. It rises steeply through $0.5$ at $\nbar\simeq1.4$,
and does so at the same $\nbar$ for every $N$ from $5$ to $20$ --- the boundary is set
by the photon number, not by $\nbar/N$. Beyond it the $20$-period envelope carries
several rhythms of comparable weight; at $N=50$, $\nbar=2$ these are the components near
$83$ and $91\,\gamma t$ noted in Sec.~\ref{sec:echolaw}. We therefore make no claim
about the echo \emph{period} for $\nbar\gtrsim1.5$ on the basis of long windows. Over
short windows the train remains clean to larger $\nbar$ --- the first four echoes at
$N=50$, $\nbar=2$ (Fig.~\ref{fig:echo}) are spaced $85.7/g$ apart, and
Figs.~\ref{fig:echo}, \ref{fig:acrossN}, \ref{fig:dist} and \ref{fig:detune} are all at
$\nbar=2$ --- but a four-echo
measurement cannot resolve the $2\%$ by which the two laws differ there (that spacing
lies below both), so the figures display the echo without adjudicating between the
forms. The adjudication is Fig.~\ref{fig:grid}(a).

\paragraph{Between the echoes.} The intensity does not simply vanish between recurrences:
it settles at
\begin{equation}
S \;\longrightarrow\; \frac{\nbar}{2},
\label{eq:collapse}
\end{equation}
i.e.\ half the available photons absorbed. This is approached as $N$ grows --- within
$1.4\%$ at $N=20$, $0.5\%$ at $N=50$, and $0.1\%$ at $N=400$ --- and holds equally at
$\nbar=5$ and $\nbar=10$. Equation~\eqref{eq:collapse} is not an independent result: once
the contributing blocks dephase, each $\sin^{2}$ term in Eq.~\eqref{eq:S} averages to
$\tfrac12$, so the collapse level is the time-average of the same expression that
produces the echoes. The collapse is not perfectly quiet, however: a residual
oscillation of order $6\%$ of the full amplitude persists at $N=50$, $\nbar=2$,
decreasing with $N$ (to $3\%$ at $N=400$) and increasing with $\nbar$ (to $15\%$ at
$\nbar=10$). The collapse is therefore cleanest in precisely the regime of interest here,
many molecules and few photons.

At long times the initial echo set fades --- on a scale $\sqrt{N}\,\tau_E$, or
$7.1\,\tau_E$ at $N=50$ --- and a complete set then re-emerges at
\begin{equation}
\tau_{\rm re}=\frac{\pi N}{4}\,\tau_E
\label{eq:reemerge}
\end{equation}
and multiples thereof \cite{Tavis2017}. At $N=50$ this predicts $39.3\,\tau_E$. We
measure the re-emergence at $\tau\simeq40,78,118,158$ --- a spacing of $39.3$ --- and
obtain $38.1$, $39.2$ and $39.2$ for coherent, smooth-squeezed and oscillatory-squeezed
distributions respectively (Sec.~\ref{sec:dist}). Equation~\eqref{eq:reemerge} is
therefore confirmed to $\sim3\%$ across three distinct photon distributions.
Figure~\ref{fig:echo}(b) shows the echoes at $\tau=35$--$45$, still at unit spacing and
full contrast as the first re-emergence builds. This confirms and extends the
re-emergence reported in Ref.~\cite{Tavis2017}; the extended arXiv version of that work
\cite{Tavis2017arXiv} treats the associated entropy, entanglement and $Q$-function
behaviour, which we do not revisit here.

\section{Independence of the initial photon distribution}
\label{sec:dist}

\begin{figure*}[!tb]
\centering
\begin{minipage}{0.48\textwidth}\centering
\includegraphics[width=\textwidth]{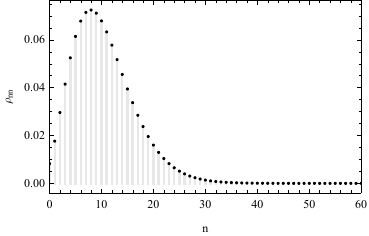}\\{\small (a) smooth}
\end{minipage}\hfill
\begin{minipage}{0.48\textwidth}\centering
\includegraphics[width=\textwidth]{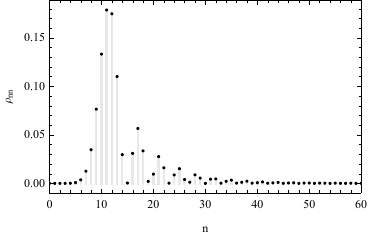}\\{\small (b) oscillatory}
\end{minipage}\\[8pt]
\includegraphics[width=\textwidth]{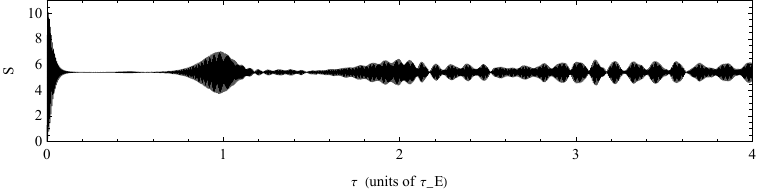}\\[-1pt]{\small (c) $S(t)$, smooth}\\[5pt]
\includegraphics[width=\textwidth]{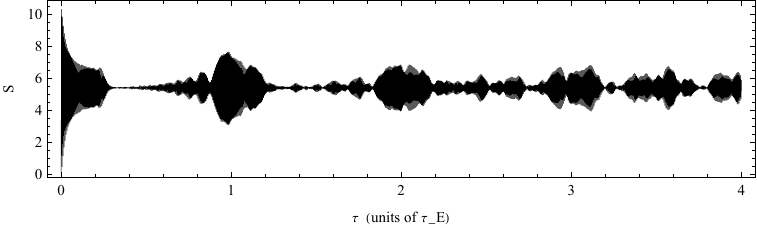}\\[-1pt]{\small (d) $S(t)$, oscillatory}
\caption{Distribution independence: a controlled pair. Two squeezed coherent states with
\emph{matched} mean and variance, differing only in the shape of $\rho_{nn}$.
(a) $r=0.6$, $\phi=0$, $\nbar_c=10$: smooth and unimodal, $n_{\rm mean}=10.41$,
$\mathrm{Var}=34.34$.
(b) $r=1.4$, $\phi=\pi$, $\nbar_c=6.78$: strongly oscillatory, with near-zeros at
$n\simeq10,15,22$, yet $n_{\rm mean}=10.41$ and $\mathrm{Var}=33.97$. Both panels share
the horizontal range.
(c), (d) The resulting $S(t)$ for $N=50$ ground-state molecules on resonance, on a
common vertical scale and against $\tau=t/\tau_E^{\rm plot}$ with the normalization
$\tau_E^{\rm plot}=4\pi\sqrt{N-\nbar}/g\simeq79.07/g$ at $\nbar=10.41$
(Eq.~\eqref{eq:echo} gives $88.86/g$; the comparison here is between the two panels at
identical normalization, so the choice is immaterial). The echo occurs at the same $\tau$ in both
cases --- timing is shape-blind --- but with visibly different amplitude: first-echo
swing $3.30$ (smooth) versus $4.59$ (oscillatory), the $39\%$ shape sensitivity of
Table~\ref{tab:dist}. Since mean and variance are held fixed, that difference isolates
the effect of distribution \emph{shape} on the echo amplitude, while the recurrence time
is unchanged across every distribution examined.}
\label{fig:dist}
\end{figure*}

The most experimentally consequential property of the echo is that it does not care
what the photon distribution is. This is, at bottom, a corollary of the block structure:
$S(t)=\sum_n\rho_n S_n(t)$ exactly, and the Appendix shows that the per-block frequency
depends on $n$ only at relative order $1/N$ while the per-block amplitude is
proportional to $n$ --- so at fixed $\nbar$ the echo \emph{time} cannot depend on the
shape of $\rho_n$ to leading order, while the amplitude may acquire $O(\nbar/N)$
corrections. What the numerics establish (Table~\ref{tab:dist} and Fig.~\ref{fig:dist}) is a sharp
split: the echo \emph{time} is distribution-blind to high accuracy, while the echo
\emph{amplitude} carries the subleading corrections in full --- it varies at the
tens-of-percent level across Table~\ref{tab:dist}, exactly the $O(\nbar/N)$ sensitivity
the leading-order argument permits. The
dependence of collective emission on photon statistics was examined at the ladder
endpoints in Refs.~\cite{TC2013,Tavis2015}; we extend that question to the few-photon
echo regime and to nonclassical distributions.

\begin{table*}[!tb]
\centering
\begin{tabular}{lcccc}
\toprule
distribution & Var & shape & echo time & first-echo amplitude\\
\midrule
coherent, $\nbar=2$ & 2 & smooth & $\tau=1$ & 1.88\\
thermal, $\nbar=2$ & 6 & smooth & $\tau=1$ & 1.52\\
squeezed coherent, $r=0.6$, $\phi=\pi$ & 4 & smooth & $\tau=1$ & 5.91\\
squeezed coherent, $r=0.6$, $\phi=0$ & 34 & smooth & $\tau=1$ & 3.30\\
squeezed coherent, $r=1.4$, $\phi=\pi$ & 33 & \textbf{oscillatory} & $\tau=1$ & 4.59\\
\bottomrule
\end{tabular}
\caption{Distribution independence at $N=50$, ground state, resonance, at fixed mean
within each comparison. Variance spans a factor of $17$ overall ($2$ to $34$); at fixed
mean the spans are $3$ ($\nbar=2$, coherent vs thermal) and $8.5$ ($\nbar=10.4$,
squeezed pair). First-echo amplitudes are measured in the window $\tau\in[0.7,1.3]$; the global
peak-to-peak is pinned at $\nbar$ for every distribution and carries no shape
information (see text).}
\label{tab:dist}
\end{table*}

The last two rows of Table~\ref{tab:dist} are a controlled pair: identical mean photon
number ($10.41$) and matched variance ($34.3$ versus $34.0$), differing only in the
\emph{shape} of $\rho_{nn}$ --- one smooth and unimodal, the other strongly oscillatory
with near-zeros at $n\simeq10,15,22$. Any difference between these two isolates shape
alone. The echo \emph{time} shows none: both recur at $\tau=1$. The echo
\emph{amplitude} does: $3.30$ (smooth) against $4.59$ (oscillatory), a $39\%$ shape
sensitivity at matched moments. Comparing instead the two squeezing phases at fixed
$r=0.6$ (rows four and five) gives $5.91$ against $3.30$ --- a factor of $1.8$ with the
squeezing phase alone. The long-time re-emergence \emph{time} is likewise unaffected,
occurring at $\tau\simeq38$--$39$ for coherent, smooth-squeezed and oscillatory-squeezed
alike.

A caution on an earlier reading of these data: the \emph{global} peak-to-peak of $S$
(quoted in the Table~\ref{tab:dist} caption of a previous version as evidence of
amplitude independence) is dominated by the $t\approx0$ transient, which is pinned at
$\nbar$ for every distribution --- it is the total photon budget, and the same quantity
used as the convergence ceiling --- and is therefore structurally incapable of
detecting distribution dependence. The first-echo amplitudes in Table~\ref{tab:dist}
are measured in the window $\tau\in[0.7,1.3]$ and do detect it. The envelope
\emph{between} echoes retains distribution memory as well (point-by-point correlation
$0.68$ between the controlled pair).

This should be distinguished carefully from the single-atom Jaynes--Cummings
literature, where oscillatory photon distributions are long known to produce echoes
\emph{inside} revivals through phase-space interference
\cite{Satyanarayana1989,Fleischhauer1993}. The present claim is the converse and
collective: the few-photon collective echo persists \emph{regardless} of distribution.

\paragraph{Experimental consequence.} The echo \emph{timing} test requires only a noise
source: thermal microwave noise is trivially generated and injected, and the recurrence
time is unchanged by it. The amplitude sensitivity upgrades the distribution test from a
null measurement to one with a predicted signal --- the thermal echo arrives on time but
$19\%$ weaker than the coherent one at equal mean --- and the factor-of-$1.8$
squeezing-phase dependence offers a second, larger predicted signal for experiments with
a parametric amplifier. Squeezing phase controls the echo's \emph{size} while leaving
its \emph{clock} untouched.

\section{Detuning: two clean regimes with a fragmented crossover}
\label{sec:detune}

Detuning does not simply rescale the echo. Table~\ref{tab:detune} gives the measured
recurrence and amplitude at $N=50$, $\nbar=2$, all-down, as $\beta=\delta/g$ is swept.

\begin{table*}[!tb]
\centering
\begin{tabular}{crccl}
\toprule
$\beta$ & $\Delta$ & recurrence ($\tau$) & amplitude & regime\\
\midrule
0   & 0    & $0.99,1.97,2.96,3.95$ & 1.999 & resonant, clean\\
2   & 1    & irregular             & 1.958 & entering crossover\\
4   & 4    & irregular             & 1.846 & fragmented\\
6   & 9    & irregular             & 1.684 & fragmented\\
10  & 25   & irregular             & 1.321 & fragmented\\
20  & 100  & $0.687$ (spacing)     & 0.660 & dispersive\\
40  & 400  & $0.554$               & 0.221 & dispersive\\
60  & 900  & $0.525$               & 0.105 & dispersive\\
100 & 2500 & $0.507$               & 0.039 & dispersive\\
\bottomrule
\end{tabular}
\caption{Two detuning regimes. Recurrence in units of the plotting normalization
$4\pi\sqrt{N-\nbar+\Delta}$; on Eq.~\eqref{eq:echo} the unit is
$4\pi\sqrt{N+\Delta}$, larger by $2.0\%$ at $\beta=0$ and by $0.2\%$ at $\beta=40$,
since $\Delta$ dominates $\nbar$ once the detuning is appreciable. The approach of the
dispersive entries to $0.5$ is therefore unaffected.}
\label{tab:detune}
\end{table*}

Three regions are evident. For $\delta\lesssim2g$ the echo train is clean and
Eq.~\eqref{eq:echo} holds. For $\delta\sim4g$--$10g$ the train fragments: the envelope
acquires several components of comparable weight and no single recurrence exists. For
$\delta\gtrsim20g$ a clean train returns --- visibly \emph{more} regular than at
resonance --- with the recurrence converging to exactly half the resonant value:
\begin{equation}
\tau_E\;\longrightarrow\;\frac{2\pi}{g}\sqrt{N+\Delta}\;\longrightarrow\;
\frac{\pi\delta}{g^{2}}
\qquad (\delta\gtrsim20g).
\label{eq:disp}
\end{equation}
Convergence onto Eq.~\eqref{eq:disp} is $38\%$ at $\beta=20$, $11\%$ at $\beta=40$ and
$5\%$ at $\beta=60$ on the fragmented-crossover estimator used for
Table~\ref{tab:detune}; the dedicated measurement of Appendix~\ref{app:disp} brings this
to $1.6\%$ at $\beta=80$ and $0.098\%$ at $\beta=320$ for $N=5$. That the second branch
is the dispersive limit is confirmed independently by the amplitude, which falls as
$\delta^{-2}$ --- the $(g/\delta)^{2}$ suppression of excitation exchange.

\begin{figure*}[!tb]
\centering
\includegraphics[width=\textwidth]{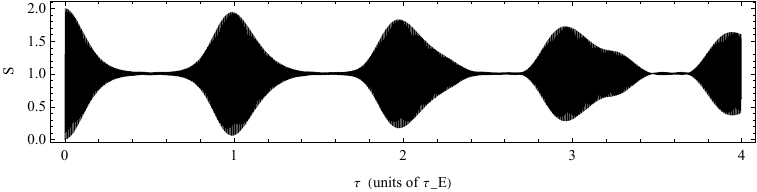}\\[-1pt]{\small (a) $\beta=0$ (resonant)}\\[4pt]
\includegraphics[width=\textwidth]{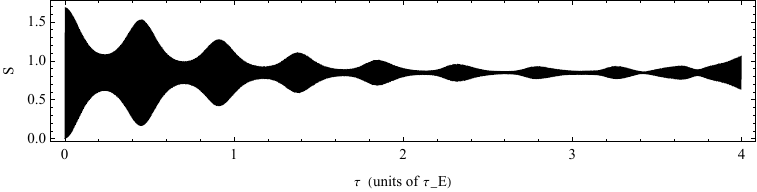}\\[-1pt]{\small (b) $\beta=6$ (crossover)}\\[4pt]
\includegraphics[width=\textwidth]{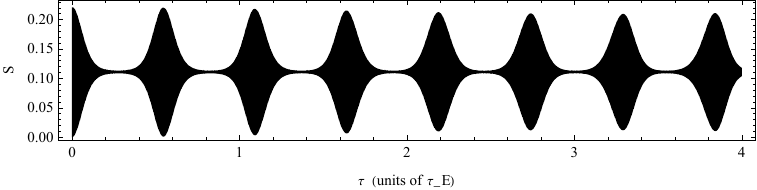}\\[-1pt]{\small (c) $\beta=40$ (dispersive)}\\[4pt]
\includegraphics[width=\textwidth]{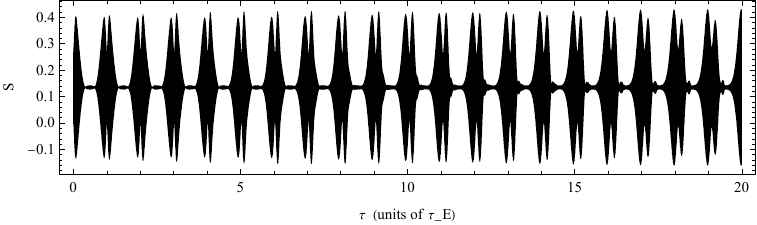}\\[-1pt]{\small (d) $\beta=40$, $m=0$, $N=20$}
\caption{The two detuning regimes and the fragmented crossover. $N=50$, ground state,
coherent field with $\nbar=2$; $\beta=\delta/g$. Each panel is plotted against
$\tau=t/\tau_E^{\rm plot}$ with the normalization
$\tau_E^{\rm plot}=4\pi\sqrt{N-\nbar+\Delta}/g$ evaluated for that detuning
($87.06$, $94.87$ and $265.98$ in units of $1/g$); on Eq.~\eqref{eq:echo} these become
$88.86$, $96.52$ and $266.57$, differing by $2.1\%$, $1.7\%$ and $0.2\%$ respectively.
(a) On resonance, four echoes at integer $\tau$.
(b) At $\beta=6$ the train fragments: the envelope carries several components of
comparable weight and no single recurrence exists.
(c) At $\beta=40$ a clean train returns --- more regular than at resonance --- with
spacing $0.554$ on this axis, approaching the one-half of Eq.~\eqref{eq:disp}.
\textbf{Note the differing vertical scales.} Peak-to-peak values are $1.999$, $1.684$
and $0.221$ respectively: the dispersive branch retains only $11\%$ of the resonant
signal, the $(g/\delta)^{2}$ suppression that identifies it as the dispersive limit.
A shared scale would render panel (c) invisible.
(d) The same dispersive branch started from the \emph{superradiant} state $m=0$
($N=20$, $\nbar=2$, $\beta=40$, shown over twenty periods): the train is equally regular.
See Sec.~\ref{sec:mindep}.}
\label{fig:detune}
\end{figure*}

Equation~\eqref{eq:echo} is derived in the Appendix, where the resonant block is shown
to be exactly harmonic. The factor of two in Eq.~\eqref{eq:disp} follows from the loss
of that harmonicity, and is derived in Appendix~\ref{app:disp} by second-order
dispersive perturbation theory: the block-to-block frequency spacing in the dispersive
limit is exactly $2g/\beta$ --- twice the $g/\beta$ implied by extrapolating the
resonant law to $\Delta\gg N$ --- so $\tau_E^{(D)}=\pi|\delta|/g^{2}$ against
$\tau_E^{(R,\rm ext)}=2\pi|\delta|/g^{2}$. We are indebted to an anonymous
referee for supplying that derivation.

In the dispersive branch the echo time is \emph{linear} in detuning and the train is
regular enough to time precisely, so this branch --- not the resonant one --- offers a
swept, falsifiable curve. The cost is amplitude: at $\beta=20$ and $40$ the echo retains
$33\%$ and $11\%$ of its resonant value, which bounds the useful range.

\subsection*{Detuning removes the dependence on the initial Dicke state}
\label{sec:mindep}

The dispersive branch has a further property that connects this section to
Sec.~\ref{sec:ladder}. Deep in that branch the recurrence no longer depends on which
Dicke state the ensemble started in. At $N=20$, $\nbar=2$, $\beta=40$ we measure
$125.7\,\gamma t$ from the superradiant state $m=0$ and $132.3\,\gamma t$ from the ground
state $m=-N/2$ --- a $5\%$ spread --- against
$2\pi\sqrt{\nbar+\Delta}=126.0$ and $2\pi\sqrt{N+\Delta}=128.8$, which themselves
differ by only $2\%$ at $\Delta=400$. Both approach the asymptote
$\pi\delta/g^{2}=125.7$.

The reason is transparent: once $\Delta\gg N$ the detuning dominates both $\nbar$ and
$N$ under the square root, and the initial collective state drops out.
Fig.~\ref{fig:detune}(d) shows the consequence directly --- at $m=0$, where the resonant
dynamics is irregular (Sec.~\ref{sec:ladder}), the detuned train is as clean as from the
ground state.

The dependence on $m$ established in Sec.~\ref{sec:ladder} is therefore a
\emph{resonant} phenomenon. The collective character of the initial state governs the
dynamics when molecules and field are resonant, and ceases to matter when they are not.
Amplitudes still differ strongly --- $0.59$ at $m=0$ against $0.095$ at $m=-N/2$, the
excitation budget of Sec.~\ref{sec:ladder} --- so detuning equalizes the \emph{timing}
without equalizing the \emph{signal}.

\section{The Dicke ladder: arbitrary initial $m$}
\label{sec:ladder}

Because Eq.~\eqref{eq:S} holds for every $m$, the full ladder is accessible at no extra
cost. We report three results at $N=50$, $\nbar=2$, resonance, with the time axis pinned
to the ground-state clock so that all $m$ are directly comparable.

\begin{figure*}[!tb]
\centering
\includegraphics[width=0.52\textwidth]{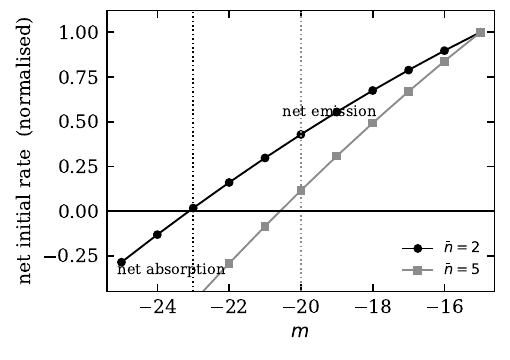}\\[2pt]{\small (a)}\\[8pt]
\includegraphics[width=\textwidth]{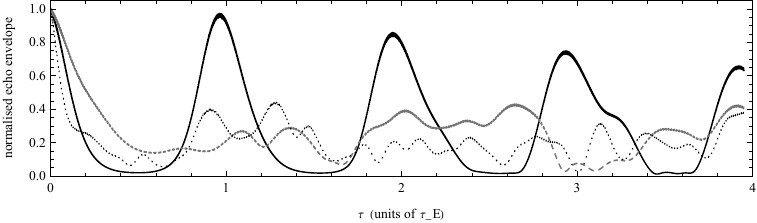}\\[-1pt]{\small (b)}\\[6pt]
\includegraphics[width=\textwidth]{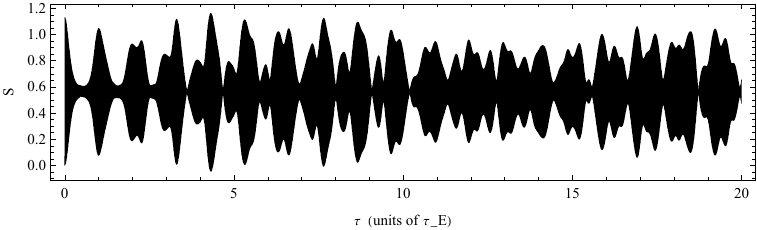}\\[-1pt]{\small (c)}
\caption{The Dicke ladder at resonance, $N=50$.
(a) Net initial rate against $m$, from
$\langle (n+1)(r+m)(r-m+1)-n(r-m)(r+m+1)\rangle_{\rho_n}$, normalised, for $\nbar=2$
(circles) and $\nbar=5$ (squares). The rate changes sign at $m_{\rm c}=-N/2+\nbar$
(dotted verticals), i.e.\ at $m=-23$ and $m=-20$ respectively --- here $m_{\rm c}$
denotes the first integer $m$ with net emission --- the ensemble ceasing to absorb once
the initially excited molecules outnumber the available photons. This is Eq.~\eqref{eq:mc}, obtained analytically and confirmed against the full
dynamics.
(b) Normalised envelope of $S(t)$ at $m=-N/2$ (solid), $m=-15$ (dashed) and $m=0$
(dotted), with a coherent field, $\nbar=2$, all on the common ground-state clock
$\tau_E^{\rm plot}=4\pi\sqrt{N-\nbar}/g=87.06/g$ so that the three are directly
comparable ($88.86/g$ on Eq.~\eqref{eq:echo}; a common normalization is what matters
here). Only the ground
state produces a clean train at integer $\tau$; the other two are irregular. The curves
are normalised individually because the amplitudes differ greatly --- $2.0$, $7.9$ and
$22.4$ respectively (the $m=0$ value requires fine sampling of the initial transient;
coarser grids clip it to $\sim19$) --- and a common scale would render the first
invisible.
(c) $S(t)$ at $m=-24$ over twenty periods on the same clock: the echo survives one step
up the ladder --- the first recurrences at $\tau=1,2$ are clean --- but the envelope now
carries two components $5\%$ apart (Table~\ref{tab:ladderfine}) whose beating blurs the
train beyond $\tau\sim10$.}
\label{fig:ladder}
\end{figure*}

\paragraph{Emission replaces absorption at a sharp value of $m$
(Fig.~\ref{fig:ladder}a).} As $m$ rises from
$-N/2$ the ensemble ceases to absorb and begins to emit, at
\begin{equation}
m_{\rm c}\simeq-\frac{N}{2}+\nbar ,
\label{eq:mc}
\end{equation}
i.e.\ when the number of initially excited molecules exceeds the photons available to
absorb. We verify $m_{\rm c}=-23$ at $\nbar=2$ and $-20$ at $\nbar=5$ exactly, and to
$\pm1$ at $\nbar=8$. At $m=m_{\rm c}$ the intensity straddles zero.

\paragraph{Amplitude is set by the excitation budget.} Peak-to-peak $S$ runs
$2.00,2.99,7.95,12.86,17.70,22.45$ for $m=-25,-20,-15,-10,-5,0$, closely tracking
$(m+N/2)-2$. This is energy bookkeeping, \emph{not} cooperativity: the $N^{2}$-type
collective enhancement appears in the initial \emph{rate}, $\propto r(r+1)-m(m-1)$, and
not in the total swing. We state this explicitly to avoid the stronger claim being read
into it.

\paragraph{The echo along the ladder: one envelope component per step.} The ground
state shows a single clean recurrence. Climbing the ladder does not extinguish the echo
--- it multiplies its envelope components. Decomposing the envelope over a
$20\,\tau_E$ window gives, at $N=50$, $\nbar=2$ (leading components, $\tau_E$ units,
relative weight in parentheses):

\begin{table}[t]
\centering
\small
\begin{tabular}{lll}
\toprule
$m$ & components & amplitude\\
\midrule
$-25$ & $1.05\,(1.00)$ + half-harmonics$^{\rm a}$ & 2.00\\
$-24$ & $1.11\,(1.00)$, $1.05\,(0.95)$ & 1.32\\
$-23$ & $1.05\,(1.00)$, $1.11\,(0.62)$, $0.54\,(0.61)$ & 1.38\\
$-22$ & $0.87$, $1.11$, $0.91$, $1.05$ (all $\geq0.81$) & 1.63\\
\bottomrule
\end{tabular}
\caption{Envelope decomposition at the bottom of the ladder. One component is added
per step: the count is $m+N/2+1$, the number of transition families the initial state
couples to. Components at half the period of a stronger component are counted as
harmonics, not fundamentals; the $0.54$ entry at $m=-23$ lies at the edge of that
distinction at the $5\%$ bin resolution, and the mechanism (three families at three
interior positions) is what fixes its assignment. $^{\rm a}$The $m=-25$ dominant
component ($1.05\,\tau_E$) and the $0.997\,\tau_E$ of Sec.~III are adjacent Fourier
bins of the $20\,\tau_E$ window, consistent at the stated resolution. Over much longer
windows the $m=-25$ envelope resolves fine components at $83.1$, $85.5$ and
$42.8\,\gamma t$.}
\label{tab:ladderfine}
\end{table}

The pattern is one additional component per step up the ladder, and the mechanism is
visible in the Appendix picture: at $m=-N/2$ the initial state is the \emph{end} vector
of each block and couples to a single transition family; at $m=-N/2+1$ it sits one
position interior and couples to two families whose frequencies differ at relative order
$1/N$; and so on. Over a few periods the echo is therefore essentially intact one or two
steps up --- at $m=-24$ the first recurrences at $\tau=1,2$ are clean, and the two
components ($5\%$ apart) only blur the train after $\sim10$ periods --- while higher on
the ladder the accumulating components dephase the envelope immediately. At $m=-15$ and
$m=0$ no stable single period remains over windows from $1200$ to $6000\,\gamma t$
(long-window decompositions: $199.5$--$213.8$ vs $101.4\,\gamma t$ at $m=-15$; $85.5$,
$110.8$, $61.1\,\gamma t$ at $m=0$). Only the ground state is single-component.
Peak-to-peak amplitudes at the bottom of the ladder ($1.32$, $1.38$, $1.63$ for
$m=-24,-23,-22$) follow neither $(m+N/2)$ nor $|m+N/2-\nbar|$ in any simple way and are
quoted as measured.

We note that this $m$ dependence is confined to resonance --- Sec.~\ref{sec:mindep}
shows that sufficient detuning removes it entirely --- and that the ground state being
the sole single-component case is experimentally convenient: it is the one initial state
obtained for free.

\paragraph{The vacuum limit shows the mechanism directly.} With no photons present the
dynamics is confined to a \emph{single} block, of dimension
\begin{equation}
d(m)=m+\frac{N}{2}+1 ,
\label{eq:dim}
\end{equation}
so the number of contributing frequencies is fixed by $m$ alone and no averaging over
$\rho_n$ intervenes. The character of the motion then follows immediately
($N=50$, resonance):

\begin{center}
\small
\begin{tabular}{rrrrl}
\toprule
$m$ & $d(m)$ & freq. & ampl. & behaviour\\
\midrule
$-25$ & 1 & 0 & 0.00 & eigenstate\\
$-24$ & 2 & 1 & 1.00 & one Rabi frequency\\
$-23$ & 3 & 3 & 2.00 & beating\\
$-20$ & 6 & 15 & 5.00 & collapse, revival\\
$-10$ & 16 & 120 & 14.97 & collapse, revival\\
$0$ & 26 & 325 & 24.88 & collapse, revival\\
$+25$ & 51 & 1275 & 40.98 & collapse, revival\\
\bottomrule
\end{tabular}
\end{center}

At $m=-N/2$ the block is one-dimensional and nothing evolves; at $m=-N/2+1$ it is
two-dimensional and the motion is a single undamped oscillation; thereafter the number of
pairwise frequencies grows as $d^{2}$ and their mutual dephasing produces collapse and
revival. The $m$ dependence of Sec.~\ref{sec:ladder} is therefore, at bottom, a statement
about how many collective frequencies participate. With photons present this mechanism is
compounded by dephasing \emph{between} blocks; at vacuum it acts alone.

Two checks follow. For $m\leq0$ the amplitude equals $m+N/2$, the number of initially
excited molecules --- $1.00$, $5.00$, $14.97$, $24.88$ against $1$, $5$, $15$, $25$.
Above the middle of the ladder the equality degrades over finite windows: at $m=+25$ the
tabulated $40.98$ falls $18\%$ short of $50$, the recurrence of the largest block being
slower than the window examined. And Eq.~\eqref{eq:dim} gives $d(0)=26$ at $N=50$: twenty-six collective frequencies
for the superradiant vacuum state.

It is worth noting the structural parallel with Sec.~\ref{sec:detune}: in both cases the
dynamics is clean in two limits and fragmented in between, and in both the two clean
limits are related by a simple factor --- exactly $2$ for detuning, exactly $1$ for the
ladder. Both point the same way: the collective echo is a property of states of definite
collective character, and mixing destroys it. We offer this as an observation rather than
an explanation.

\section{Experimental realization}
\label{sec:expt}

\subsection{Benchmark device and scope}
\label{sec:bench}

The initial state is the ground state; no Dicke-state engineering, ensemble $\pi$-pulses
or nonclassical source is required. The closest existing experiment is the
superconducting atomic frequency comb of Ref.~\cite{Redchenko2025}, which realizes the
periodic cavity-state revivals predicted in Ref.~\cite{Zens2021} using an
\emph{inhomogeneous} ensemble; the echo proposed here is the homogeneous, few-photon
counterpart and requires no comb engineering. We benchmark against the measured parameters of an
existing superconducting device \cite{Redchenko2025}: five transmons coupled to a
half-wave coplanar-waveguide resonator at $\omega_c/2\pi=5.878\,$GHz, single-qubit
couplings $g_k/2\pi=28.1$--$32.3\,$MHz, collective $G/2\pi=68.95\,$MHz, loaded linewidth
$\kappa_{\rm load}/2\pi=0.93\,$MHz, \emph{internal} loss
$\kappa_i/2\pi\approx3\,$kHz, qubit decoherence $\gamma_k/2\pi$ from $<10\,$kHz to
$470\,$kHz, revival dynamics reported intact over a drive-power range calibrated as $0.1$--$38$
input photons, and qubit
frequencies flux-tunable over $>2\,$GHz. With $g/2\pi=30\,$MHz, $1/g=5.31\,$ns.

\begin{table}[h]
\centering
\begin{tabular}{ccrrrr}
\toprule
$N$ & $\nbar$ & $4\pi\sqrt{N}$ & $\tau_E$ & $\times\kappa_{\rm load}$ &
$\times\kappa_i$\\
\midrule
5  & 1 & 28.10 & 149 ns & 0.87 & 0.0028\\
10 & 2 & 39.74 & 211 ns & 1.23 & 0.0040\\
20 & 2 & 56.20 & 298 ns & 1.74 & 0.0056\\
50 & 2 & 88.86 & 471 ns & 2.75 & 0.0089\\
\bottomrule
\end{tabular}
\caption{Predicted echo times against the loss budget of Ref.~\cite{Redchenko2025},
computed from Eq.~\eqref{eq:echo} throughout. The $N=5$ entry agrees with the directly
measured recurrence $27.8\,\gamma t$ (Sec.~\ref{sec:echo}) to $1\%$. An earlier version
of this table used the alternative normalization $4\pi\sqrt{N-\nbar}$, which shortens
every entry by $1$--$12\%$; the conclusions are unchanged in substance.}
\label{tab:feas}
\end{table}

\begin{figure}[t]
\centering
\includegraphics[width=\columnwidth]{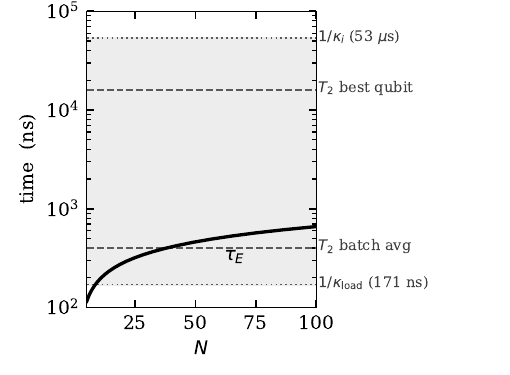}
\caption{Feasibility against the measured parameters of Ref.~\cite{Redchenko2025}.
Solid curve: the predicted echo time $\tau_E=4\pi\sqrt{N}/g$ of Eq.~\eqref{eq:echo} with
$g/2\pi=30\,$MHz, rising from $149\,$ns at $N=5$ to $667\,$ns at $N=100$. (The curve as
drawn uses the earlier normalization $4\pi\sqrt{N-\nbar}/g$ at $\nbar=2$, which runs
from $115\,$ns to $660\,$ns; the difference is $23\%$ at $N=5$, falling to $1\%$ by
$N=100$, and moves no curve across a threshold line. Numerical statements in the text
and in Table~\ref{tab:feas} use Eq.~\eqref{eq:echo}.) Horizontal
lines mark the loaded cavity lifetime $1/\kappa_{\rm load}=171\,$ns, the batch-average
qubit coherence $T_2\approx398\,$ns, the best qubit on the same chip
($T_2\approx16\,\mu$s), and the \emph{intrinsic} cavity lifetime
$1/\kappa_i=53\,\mu$s. The shaded band is the margin recoverable by reducing the
external coupling: $\kappa_{\rm load}$ exceeds $\kappa_i$ by a factor of $300$ because
the resonator is deliberately over-coupled for readout speed, not because of any
physical limit. Against $\kappa_{\rm load}$ and the batch-average $T_2$ the echo is
marginal beyond $N\simeq10$ and $N\simeq40$ respectively; against the intrinsic loss and the best available qubits it sits one to two orders of
magnitude inside the budget throughout --- a ceiling rather than an operating point
(see text). The feasibility claims of Sec.~VII are scoped to $N\approx5$--$20$.}
\label{fig:feas}
\end{figure}

The first row of Table~\ref{tab:feas} is now backed by direct simulation: at $N=5$,
$\nbar=1$ the echo exists but is ragged, with a dominant recurrence of $27.8\,\gamma t$
(Sec.~III) rather than a clean train --- consistent with the small-$N$ end of an
asymptotic law, and to be read accordingly. Figure~\ref{fig:feas} summarizes the loss
budget. On the device as configured the echo lands at $0.87$--$2.75$ loaded cavity
lifetimes. The intrinsic-loss column of Table~\ref{tab:feas} bounds the \emph{ceiling},
not an operating point: injecting and detecting the field requires finite external
coupling, whose optimum is $\kappa_{\rm ext}\sim1/\tau_E$, so the echo is
intrinsically a $\kappa\tau_E\sim1$ measurement --- which the cavity-decay simulation
below shows it survives with $\gtrsim50\%$ contrast. The qubit budget at
$N\leq20$ is comfortable even at the batch-average
$\gamma/2\pi\approx400\,$kHz, and improves tenfold with the best qubit on the same
chip ($\gamma/2\pi<10\,$kHz). \emph{At this scale the requirement is a
re-optimization of an existing platform, not a new one.}

We scope the feasibility claims to $N\approx5$--$20$. Beyond that, three
qualitatively new obstacles appear which no existing device has faced: at $N=50$ the
collective coupling $g\sqrt{N}/2\pi\approx212\,$MHz becomes comparable to transmon
anharmonicity, straining the two-level description; stray qubit--qubit couplings break
the symmetric-sector structure in ways the disorder model below does not capture; and a
$50\times50$ flux-crosstalk calibration is a campaign in itself (the $7\times7$ matrix
was already a supplement-level effort in Ref.~\cite{Redchenko2025}). At $N\leq20$ none
of these arises, and --- as shown below --- every simulated decoherence channel is
survivable there on demonstrated hardware.

\subsection{Loss channels individually}
\label{sec:channels}

The loss channels have been checked by direct simulation.

\paragraph{Disorder.} The block structure assumes identical, degenerate
molecules, while the benchmark device has $g_k/2\pi=28.1$--$32.3\,$MHz ($\pm7\%$) and
$\sim1\,$MHz frequency-positioning precision. We therefore simulated the echo with the
symmetry actually broken: the full $2^N$ spin space (no permutation symmetry assumed)
at $N=6$, $\nbar=2$, with Gaussian frequency disorder of spread $\sigma$ and with the
device's coupling spread. The echo is far more robust than a naive dephasing estimate
($\sigma\tau_E\lesssim1$) would suggest: contrast is degraded not on the scale
$1/\tau_E$ but on the scale of the \emph{collective splitting} $g\sqrt{N}$ --- retaining
$90\%$ at $\sigma=0.05\,g\sqrt{N}$, $76\%$ at $0.24\,g\sqrt{N}$, and $58\%$ at
$0.41\,g\sqrt{N}$ --- the cavity-protection effect familiar from inhomogeneous spin
ensembles \cite{Diniz2011,Kurucz2011}. The scaling variable was tested across $N$: at fixed $\sigma/(g\sqrt{N})$ the contrast
ratio shows no systematic $N$-trend over $N=4$--$7$ (scatter $\pm0.1$ at four
realizations), while $\sigma\tau_E$ nearly doubles across the same range and would have
predicted strong degradation. The device's $\pm7\%$ coupling spread costs nothing
measurable, and its $\sim1$--$1.5\,$MHz inhomogeneity corresponds to
$\sigma/g\sqrt{N}\approx0.01$--$0.02$ at $N=5$--$20$: disorder is a non-issue at the
demonstrated homogeneity. Cavity protection has been observed directly in spin
ensembles \cite{Putz2014}.

\paragraph{Cavity decay --- the dominant channel.} A Lindblad simulation with photon loss at
rate $\kappa$ ($N=4$, $\nbar=1$) gives echo contrast $0.70$, $0.56$, $0.41$ and $0.34$
at $\kappa\tau_E=0.5$, $1$, $2$ and $2.7$ --- close to $e^{-\kappa\tau_E/2}$ rather than $e^{-\kappa\tau_E}$ (the effective rate
drifts from $0.7\kappa$ to $0.4\kappa$ over the range tested), because the excitation
spends roughly half its time absorbed in the emitters, where it is immune to cavity
loss. Even at the benchmark device's
over-coupled $\kappa_{\rm load}\tau_E\simeq0.9$ ($N=5$) more than half the contrast
survives; the echo is attenuated by the dominant channel, not extinguished.

\paragraph{Qubit dephasing.} A Lindblad simulation with individual-qubit pure dephasing at
rate $\gamma_\phi$ gives an effective echo-decay rate that \emph{grows} with $N$:
$\Gamma_{\rm eff}/\gamma_\phi\simeq0.68$, $0.79$, $1.09$, $1.37$ at $N=2$--$5$
($\nbar=1$), a fit $\Gamma_{\rm eff}\approx0.25\,N\gamma_\phi$ --- well below the full
$N\gamma_\phi$ (at $N=4$, $\gamma_\phi\tau_E=1$, the measured contrast is $0.39$
against $0.018$ under $N\gamma_\phi$ scaling, a factor of twenty), but not
$N$-independent. With the device's batch-average $\gamma_\phi/2\pi\approx400\,$kHz this costs little at
$N\approx5$ ($\Gamma_{\rm eff}\tau_E\approx0.5$, contrast $\sim0.6$); at $N=10$ it
gives $\Gamma_{\rm eff}\tau_E\approx1.2$ (contrast $\sim0.3$ --- workable but no longer
comfortable), and better-than-average qubits restore the margin ($\approx0.03$ at the
best qubit on the chip). Extrapolated to $N=50$ it would require best-tier qubits
throughout, one more reason for the $N\leq20$ scoping. Growth with $\nbar$ is
mild ($1.0\to1.2\,\gamma_\phi$ from $\nbar=1$ to $2$ at $N=4$).

\paragraph{Undercoupling.} Recovering the
intrinsic-loss margin by reducing the external coupling also reduces the collected
signal; the optimum is $\kappa_{\rm ext}\sim1/\tau_E$, within a factor of three of the
device's present coupling at $N=50$. This trade-off has not been simulated and is
quoted as an estimate.

\paragraph{Injection.} Loading through an undercoupled port takes
$\sim1/\kappa_{\rm ext}\gtrsim\tau_E$; the natural protocol is to load with the
molecules detuned, then flux-tune into resonance, which the $>2\,$GHz tuning range
permits.

\subsection{All channels acting simultaneously}
\label{sec:allnoise}

The simulations above isolate one mechanism at a time, which is the right way to
identify each effect but not the right way to predict an experiment: a real device
suffers all of them at once, and the question of how they combine cannot be settled by
inspecting them separately. We therefore repeated the calculation with every channel
switched on together --- cavity loss $\kappa$, qubit relaxation $\gamma_1$, pure
dephasing $\gamma_\phi$, static frequency disorder of spread $\sigma$, and coupling
inhomogeneity $\epsilon$ --- in a single Lindblad master equation on the full spin
space, with the disorder realized as an explicit random draw and averaged over
realizations.

The channel strengths are taken from the benchmark device \cite{Redchenko2025} and
expressed in units of $g$: $\kappa/g=0.031$, $\gamma_1/g=\gamma_\phi/g=0.0157$,
$\sigma/g=0.02$ and $\epsilon=0.04$ (the measured $\pm7\%$ coupling spread). Echo
contrast is defined as the first-echo swing relative to the same quantity with all
channels off.

\begin{table}[h]
\centering
\begin{tabular}{lr}
\toprule
channel(s) at device values & contrast\\
\midrule
cavity loss $\kappa$ alone               & 0.585\\
qubit relaxation $\gamma_1$ alone        & 0.727\\
pure dephasing $\gamma_\phi$ alone       & 0.653\\
frequency disorder $\sigma$ alone        & 0.999\\
coupling inhomogeneity $\epsilon$ alone  & 0.980\\
\midrule
naive additive rule $1-\sum_i(1-c_i)$    & $-0.055$\\
product rule $\prod_i c_i$               & 0.272\\
\textbf{all five simultaneously}         & \textbf{0.387}\\
\bottomrule
\end{tabular}
\caption{Combined-channel echo contrast at $N=4$, $\nbar=1$, resonance, with the
benchmark device's parameters. The additive rule is not merely inaccurate but
unphysical, predicting a negative contrast; the product rule is pessimistic by $30\%$.
The channels are sub-multiplicative because the dominant ones --- cavity loss and qubit
decoherence --- act on the excitation in complementary halves of its cycle.}
\label{tab:allnoise}
\end{table}

Three things follow. First, the echo survives the full device noise with $39\%$ of its
ideal contrast --- not a marginal signal, and well above the $27\%$ a product rule would
predict. Second, the two configurational channels, frequency disorder and coupling
inhomogeneity, cost almost nothing at the demonstrated homogeneity ($0.999$ and
$0.980$), confirming from the combined calculation what Sec.~\ref{sec:channels}
established separately: on this device the budget is set by $\kappa$ and by qubit
decoherence, not by sample inhomogeneity. Third, the combination is \emph{better} than
multiplicative, which is the physically meaningful statement --- the excitation spends
roughly half its time in the cavity, where it is exposed to $\kappa$, and half in the
emitters, where it is exposed to $\gamma_1$ and $\gamma_\phi$, so the two dominant
channels cannot both act at full strength at the same instant.

\begin{figure*}[!tb]
\centering
\includegraphics[width=\textwidth]{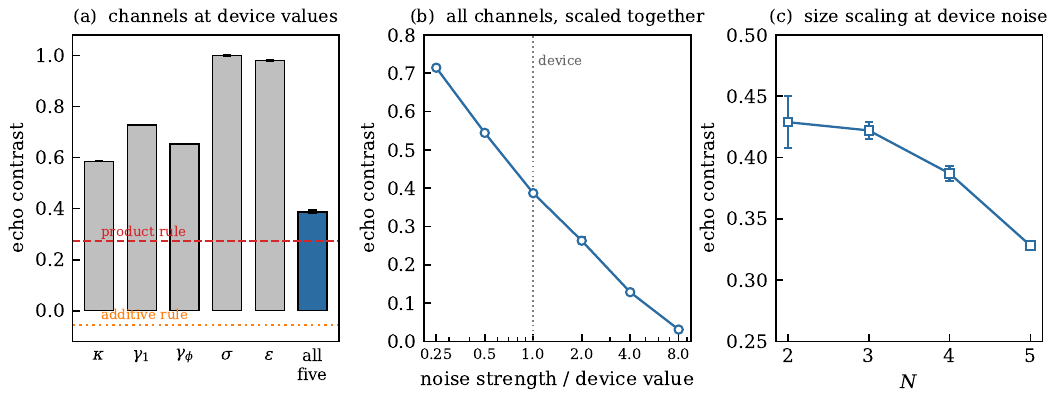}
\caption{Echo contrast under simultaneous noise channels. Lindblad simulation on the
full spin space at $\nbar=1$, resonance, with cavity loss, qubit relaxation, pure
dephasing, static frequency disorder and coupling inhomogeneity; error bars are the
spread over disorder realizations.
(a) Each channel alone at the benchmark device's values \cite{Redchenko2025} (grey) and
all five together (blue), $N=4$. The dashed and dotted lines are the product rule
$\prod_ic_i=0.272$ and the additive rule $1-\sum_i(1-c_i)=-0.055$. The measured
combined contrast, $0.387$, exceeds both: the channels are sub-multiplicative.
(b) All five scaled together by a common factor relative to the device values (dotted
vertical), $N=4$. Useful contrast survives a factor of two of across-the-board
degradation.
(c) Size scaling at fixed device noise. The decline from $N=2$ to $N=5$ is slow,
far gentler than an independent-emitter $N\gamma_\phi$ estimate would predict.}
\label{fig:allnoise}
\end{figure*}

Figure~\ref{fig:allnoise} gives the scaling. Holding the ratios fixed and scaling all
five strengths together by a common factor, contrast falls from $0.715$ at one-quarter
of the device values through $0.387$ at the device values to $0.031$ at eight times
them: useful contrast ($>0.25$) survives a factor of two of across-the-board
degradation, so the prediction does not depend on the benchmark device being
reproduced exactly. Scaling $N$ at fixed device noise gives $0.429$, $0.422$, $0.387$
and $0.328$ at $N=2,3,4,5$ --- a slow decline, consistent with the
$\Gamma_{\rm eff}\approx0.25N\gamma_\phi$ of Sec.~\ref{sec:channels} and with
$\tau_E\propto\sqrt N$ lengthening the exposure, and far gentler than the
$N\gamma_\phi$ that a naive independent-emitter estimate would give.

\subsection{What is tested and what is assumed}
\label{sec:caveats}

The simulations reported here are small --- $N\leq6$ on the full spin space --- and it
is worth separating what they establish from what remains an assumption, so that the
feasibility claim can be read at its proper strength.

\emph{Tested numerically.} The echo exists and is timed by Eq.~\eqref{eq:echo} over
$N=5$--$400$; its period is independent of $\nbar$ wherever a single recurrence exists,
which is for $\nbar\lesssim1.4$ at every $N$ examined (Sec.~\ref{sec:grid}); it is
insensitive to the
photon distribution (Sec.~\ref{sec:dist}); it survives each decoherence channel
separately (Sec.~\ref{sec:channels}) and all of them together at the benchmark device's
measured parameters, with $39\%$ contrast at $N=4$ (Sec.~\ref{sec:allnoise}); the
contrast degrades only slowly with $N$ over the range accessible to full-space
simulation; and cavity protection against disorder operates on the scale $g\sqrt N$
rather than $1/\tau_E$.

\emph{Assumed, and requiring experimental validation.} That the slow $N$-dependence of
the combined contrast observed over $N=2$--$5$ continues to $N=10$--$20$; the full spin
space grows as $2^N$ and we cannot simulate the target range directly. That the
transmons remain adequately two-level at $N\leq20$, where
$g\sqrt N/2\pi\approx95$--$134\,$MHz is already a substantial fraction of a typical
anharmonicity. That stray qubit--qubit couplings, which our disorder model does not
include, do not break the symmetric sector in a way that destroys the echo. That the
flux ramp from the loading detuning through the fragmented crossover
($\beta\sim2$--$20$) can be made fast compared with $1/(g\sqrt N)\approx1$--$2\,$ns and
synchronous across all bias lines --- the most demanding control requirement in the
protocol, and the one we regard as least certain. That the external coupling can be
re-optimized toward $\kappa_{\rm ext}\sim1/\tau_E$ without a loss of collection
efficiency we have not modelled. And that moment-inversion detection reaches the
required sensitivity at the repetition counts estimated below.

None of these is exotic, but none is established by the calculations presented here, and
we do not claim otherwise.

\subsection{Protocol}

(i) Cool; the initial state is the ground state.
(ii) Inject a weak field with $\nbar<N$, within the demonstrated $0.1$--$38$ photon range.
(iii) Record the emitted flux $\kappa_{\rm ext}\langle\ad a\rangle(t)$
\cite{GardinerCollett} at the existing
$\sim2\,$ns resolution and identify the echo at Eq.~\eqref{eq:echo}. Note the observable
throughout this paper is the photon \emph{number}, not a field quadrature: a
linear-amplification chain must reconstruct $\langle\ad a\rangle$ by per-shot squaring
with noise-power subtraction (moment inversion \cite{daSilva2010,Eichler2011}); this is
essential for the thermal-input test, where $\langle a\rangle=0$ and only the power
carries the signal. A rough budget: at $N=5$--$20$ the echo feature is a fraction of a
photon of emitted energy against an amplifier noise of $10$--$20$ photons (HEMT), so
$10^{4}$--$10^{6}$ repetitions per trace --- seconds to minutes at microsecond
repetition periods --- or $10$--$100\times$ fewer with a near-quantum-limited
amplifier.
(iv) Sweep $\delta$ by flux bias across both branches of Sec.~\ref{sec:detune}; the
dispersive branch requires $\delta$ up to $\sim1.2\,$GHz, comfortably within the
available tuning.
(v) Repeat with injected broadband noise at matched $\nbar$: the echo should arrive on
time but $19\%$ weaker (Sec.~\ref{sec:dist}) --- a predicted signal, not a null test.
Injection rather than heating: the equilibrium occupation at $10\,$mK and
$5.878\,$GHz is $n_{\rm th}\approx6\times10^{-13}$, and though measured residual
occupations in real cryostats are typically $10^{-3}$--$10^{-1}$, both are negligible
against $\nbar\sim1$. Finally, the flux ramp from the loading detuning to resonance
must traverse the fragmented crossover ($\beta\sim2$--$20$) quickly compared to
$1/(g\sqrt{N})\approx1$--$2\,$ns, synchronously across all bias lines; slower ramps
deposit the state into the crossover regime where no clean echo exists.

Step (v) is the cheapest high-information measurement in the sequence.

\paragraph{What is observable.} Explicitly: the first echoes at $\tau_E$ --- yes, at
$N\approx5$--$20$ on existing hardware, per the simulations above. The dispersive-branch
period halving --- partially: the practical window is $\beta\approx10$--$30$, since at
$\beta=40$ the echo is $780\,$ns at $0.22$ photons with qubits parked $1.2\,$GHz from
their sweet spots, and at $\beta=100$ it is $1.7\,\mu$s at $0.04$ photons. The
long-time re-emergence at $\pi N\tau_E/4$ --- no: at $N=50$ it lies at
$\approx18\,\mu$s, beyond every coherence budget in Table~\ref{tab:feas}; it is a
prediction for future hardware. The $m$-dependence of Sec.~VI requires preparing
excited Dicke states and is likewise beyond the immediate protocol.

\section{Discussion}

The results above are all consequences of a solution published in 1968, evaluated in a
regime that solution's own literature has not occupied. That regime is not exotic: it is
the regime an experiment naturally lands in, because $\nbar<N$ with the ensemble in its
ground state is the easiest configuration to prepare.

Two of the findings restrict the claim rather than extend it, and we regard that as a
feature. The detuning dependence is not a smooth square-root curve but two regimes with
an incoherent zone between them. The Dicke ladder does not support a single echo law but
rather a clean echo at each end and no stable recurrence in the middle. Both of these
were found by looking for the simpler behaviour and failing to find it.

The factor of two between the resonant and dispersive branches, left as a numerical
result in an earlier version of this work, is now derived: Appendix~\ref{app:disp} gives
it from a second-order dispersive expansion, with the leading correction
$5Ng^{2}/\delta^{2}$ verified against the full dynamics. That derivation is due to an
anonymous referee.

One question remains open. The dynamics of the ladder interior --- which has envelope
structure but no stable period --- deserves a treatment with tools better suited
to multi-component envelopes than the spectral methods used here. A second, more
practical gap is that the combined-noise simulations of Sec.~\ref{sec:allnoise} reach
only $N=5$, while the feasibility claim is made for $N\leq20$; closing that gap requires
either a symmetry-adapted open-system method or a device.

\section*{Reproducibility}

All figures are generated from the computational machinery described in
Sec.~\ref{sec:model}; the notebook and an independent cross-check implementation are
provided as supplementary material. Sampling convergence for every trace was verified
against the peak-to-peak ceiling described there.

\appendix
\section{Derivation of the echo law}
\label{app:derivation}

Equation~\eqref{eq:echo} follows from two properties of the ground-state blocks.

\paragraph{The resonant block is exactly harmonic.} For the ensemble in $|n,-r\rangle$
the block spans $|n',m'\rangle$ with $n'=0\ldots n$ and $m'=n-r-n'$, and the couplings are
\begin{equation}
V_{n'}^{2}=(n'+1)(n-n')(N-n+n'+1).
\label{eq:V2}
\end{equation}
In the few-photon regime $n\ll N$ the last factor is $N$ to leading order, so
$V_{n'}^{2}\simeq N(n'+1)(n-n')$ --- which is exactly $(2\sqrt{N})^{2}$ times the $J_x$
matrix elements of a fictitious spin $j=n/2$. Hence
\begin{equation}
H_{\rm block}\;\simeq\;2\sqrt{N}\,J_x ,
\label{eq:Jx}
\end{equation}
whose levels are \emph{equally spaced} by $2\sqrt{N}$. Numerically the spacings within a
block are uniform to $0.08\%$ at $N=50$ and $0.01\%$ at $N=200$. The block therefore
contributes no dephasing of its own: the collapse and the echo arise entirely from
dephasing \emph{between} blocks.

\paragraph{The effective coupling falls linearly in $n$.} Retaining the neglected
factor in Eq.~\eqref{eq:V2} and averaging uniformly over the block --- a heuristic whose
justification is the numerical slopes below --- gives
$V_{\rm eff}^{2}\simeq N-(n-1)/2$, so that
\begin{equation}
\frac{dV_{\rm eff}^{2}}{dn}=-\frac{1}{2}
\label{eq:dV2}
\end{equation}
independently of $N$. This is confirmed numerically: $-0.5048$ at $N=200$ and $-0.5200$
at $N=50$, approaching $-1/2$ as $N$ grows.

\paragraph{The echo time.} With detuning the dominant frequency of a two-level-like pair
is $\omega=\sqrt{\beta^{2}+4V_{\rm eff}^{2}}=2\sqrt{\Delta+V_{\rm eff}^{2}}$, using
$\Delta=\beta^{2}/4$. Then
\begin{equation}
\frac{d\omega}{dn}=\frac{2}{\omega}\frac{dV_{\rm eff}^{2}}{dn}
=-\frac{1}{2\sqrt{N+\Delta}} .
\end{equation}
Adjacent blocks rephase when their accumulated phase difference reaches $2\pi$, i.e.\
$\tau|d\omega/dn|=2\pi$, giving
\begin{equation}
\tau_E=\frac{4\pi}{g}\sqrt{N+\Delta},
\end{equation}
which is Eq.~\eqref{eq:echo}. The $\sqrt{N}$ scaling therefore originates in the
fictitious-spin mapping \eqref{eq:Jx}, and the numerical coefficient $4\pi$ in the exact
slope \eqref{eq:dV2}.

\paragraph{Why the dispersive branch differs.} The argument above uses harmonicity: with
equally spaced levels the only dephasing is between blocks. Detuning destroys that
property. At $N=50$, $n=5$, $\beta=40$ the intra-block spacings run
$42.06,\,42.20,\,42.33,\,42.47,\,42.60$ --- a spread of $1.3\%$, against uniformity at
$\beta=0$ --- and the resulting intra-block dephasing adds to the inter-block dephasing,
shortening the recurrence. This identifies the mechanism; the next subsection derives
the factor.

\section{The dispersive limit and the factor of two}
\label{app:disp}

The derivation in this appendix is due to an anonymous referee, to whom we are
indebted; we give it here with the numerical checks that fix its range of validity.

\paragraph{Second-order dispersive expansion.} In the dispersive regime
$|\delta|\gg g\sqrt{nN}$ the exchange term in Eq.~\eqref{eq:H} may be eliminated to
second order. Within the block of total excitation $n$, writing $k$ for the number of
excited molecules, the couplings of Eq.~\eqref{eq:V2} read
$V_k^{2}=(k+1)(n-k)(N-n+k+1)$ and the second-order energies are
\begin{equation}
\frac{E_{n,k}}{\hbar g}=\beta(n-k)+\frac{1}{\beta}\Big[V_{k-1}^{2}-V_{k}^{2}\Big]
+O(\beta^{-3}).
\label{eq:E2}
\end{equation}
Forming $\Omega_{n,k}=(E_{n,k}-E_{n,k-1})/\hbar$ and simplifying gives
\begin{equation}
\Omega_{n,k}=g\Big[\beta+\frac{2N-2+6k-4n}{\beta}\Big].
\label{eq:Omnk}
\end{equation}
For an ensemble prepared unexcited the initial state of the $n$-excitation block
\emph{is} the bare state $k=n$, so the relevant transition is $k=n\leftrightarrow k=n-1$
and
\begin{equation}
\Omega_n\equiv\Omega_{n,n}=g\Big[\beta+\frac{2(N+n-1)}{\beta}\Big].
\label{eq:Omn}
\end{equation}
The block-to-block spacing is therefore \emph{exactly} uniform at this order,
\begin{equation}
\Omega_{n+1}-\Omega_n=\frac{2g}{\beta},
\label{eq:spacing}
\end{equation}
and adjacent blocks rephase when $\tau\,(\Omega_{n+1}-\Omega_n)=2\pi$:
\begin{equation}
\tau_E^{(D)}=\frac{\pi\beta}{g}=\frac{\pi|\delta|}{g^{2}} .
\label{eq:tauD}
\end{equation}
Extrapolating the resonant law \eqref{eq:echo} to $\Delta\gg N$ instead gives
$4\pi\sqrt{\Delta}/g=2\pi|\delta|/g^{2}$, exactly twice
Eq.~\eqref{eq:tauD}. The factor of two in Eq.~\eqref{eq:disp} is therefore not an
empirical coincidence but the statement that the dispersive block-to-block spacing is
$2g/\beta$ where the naive extrapolation implies $g/\beta$.

\paragraph{Check 1: the algebra.} Symbolic evaluation of
$\Omega_{n,k}=(E_{n,k}-E_{n,k-1})/\hbar$ from Eq.~\eqref{eq:E2} reproduces
Eq.~\eqref{eq:Omnk} identically, and Eq.~\eqref{eq:spacing} follows with difference
exactly zero. (We note that $\Omega_n$ must be formed by substituting $k\to n$
\emph{before} shifting $n\to n+1$; performing the two substitutions in the opposite
order double-shifts the index and yields $8g/\beta$ in place of $2g/\beta$.)

\paragraph{Check 2: the transition weights.} Equation~\eqref{eq:Omn} is only relevant if
the $k=n\leftrightarrow k=n-1$ transition actually dominates $\langle\hat n\rangle(t)$.
Computing $|a_\alpha a_{\alpha'}\langle\alpha|\hat n|\alpha'\rangle|$ over the exact
block eigenbasis, the weight is overwhelmingly concentrated there: the next-largest
weight is $0.005$ (relative to $1.000$) at $N=5$, $n=3$, $\beta=40$, rising only to
$0.065$ at $N=10$, $n=4$, $\beta=20$. The splitting of that transition agrees with
Eq.~\eqref{eq:Omn} to three decimals ($40.35$ against $40.350$ at $N=5$, $n=3$,
$\beta=40$).

\paragraph{Check 3: the full dynamics.} The period predicted by Eq.~\eqref{eq:tauD} was
measured directly from $S(t)$ on an $80$-point grid spanning $N=5$--$50$,
$\nbar=0.5$--$8$ and $\beta=40$--$320$, using two independent estimators --- a linear
fit to successive envelope-maximum positions, and the dominant slow frequency of the
envelope from band-limited spectral analysis. The envelope is constructed exactly, as
the modulus of the analytic signal assembled from the spectral decomposition of $S(t)$;
this removes the fast Rabi carrier and with it the sampling error that otherwise limits
the estimate at large $\beta$. Where the two estimators agree to better than $0.5\%$ the
dispersive regime is established.

\begin{table}[h]
\centering
\begin{tabular}{crrrr}
\toprule
 & $\beta=40$ & $\beta=80$ & $\beta=160$ & $\beta=320$\\
\midrule
$N=5$  & $+1.55\%$ & $+0.39\%$ & $+0.098\%$ & $+0.025\%$\\
$N=10$ & ---       & $+0.78\%$ & $+0.196\%$ & $+0.048\%$\\
$N=20$ & ---       & $+1.56\%$ & $+0.390\%$ & $+0.098\%$\\
$N=50$ & ---       & ---       & $+0.977\%$ & $+0.244\%$\\
\bottomrule
\end{tabular}
\caption{Deviation of the measured dispersive echo period from
Eq.~\eqref{eq:tauD}, at $\nbar=1$. Entries marked --- are points at which the two
estimators disagree, i.e.\ $\beta$ is not yet large enough for the expansion at that
$N$.}
\label{tab:disp}
\end{table}

\noindent Across the grid the deviation is positive --- the measured period approaches
$\pi|\delta|/g^{2}$ from above --- and is described by
\begin{equation}
\frac{\tau_E^{\rm meas}}{\pi|\delta|/g^{2}}-1\;\simeq\;\frac{5N}{\beta^{2}}
=\frac{5Ng^{2}}{\delta^{2}},
\label{eq:dispcorr}
\end{equation}
which reproduces every in-regime entry of Table~\ref{tab:disp} to better than $1\%$ of
itself; an unconstrained power-law fit returns $N^{0.94}\beta^{-1.99}$. The dependence
on $\nbar$ is very weak: at $N=20$, $\beta=160$ the deviation moves only from $0.386\%$
to $0.444\%$ as $\nbar$ ranges over $0.5,1,2,4,8$. Equation~\eqref{eq:dispcorr} is
consistent in order with the validity condition $|\beta|\gg\sqrt{nN}$, and implies that
Eq.~\eqref{eq:tauD} is accurate to better than $1\%$ whenever
$|\delta|\gtrsim22\,g\sqrt{N}$.

The same $N/\beta^{2}$ scaling governs the purely spectral discrepancy: comparing
$\Omega_{n+1}-\Omega_n$ from exact diagonalization with the second-order value
$2g/\beta$ gives $5.5\%$, $1.4\%$ and $0.36\%$ at $N=5$ for $\beta=20,40,80$, and
$10.9\%$ and $20.3\%$ at $N=10$ and $20$ for $\beta=20$ --- falling as $\beta^{-2}$ and
growing linearly in $N$, with essentially the coefficient of Eq.~\eqref{eq:dispcorr}.
The spectral and dynamical checks therefore agree not only in conclusion but in their
leading correction.

Throughout the grid the measured period lies $49$--$50\%$ below the resonant-branch
extrapolation $2\pi|\delta|/g^{2}$.

\end{document}